\documentclass[twoside]{elsart}

\usepackage{amsmath,epsfig,amsfonts,amssymb,latexsym}

\begin{document}
\date{\today}
\begin{frontmatter}
\title			{Anomalous waiting times in high-frequency financial data}
 
\author[Alessandria]	{Enrico Scalas},
\author[Berlin]        {Rudolf Gorenflo},
\author[Bologna]		{Francesco Mainardi},
\author[Alessandria]    {Maurizio Mantelli}
and
\author[Genova]		{Marco Raberto}

\address[Alessandria]	{Dipartimento di Scienze e Tecnologie Avanzate, 
			Universit\`a del Piemonte Orientale, 
			via Cavour 84, 
			I--15100 Alessandria, Italy}
\address[Berlin] {Erstes Mathematisches Institut, Freie Universit\"at Berlin, \\
                 Arnimallee 3, D--14195 Berlin, Germany}
\address[Bologna] {Dipartimento di Fisica, Universit\`a di Bologna 
                  and INFN Sezione di Bologna, \\ 
                  via Irnerio 46, I--40126 Bologna, Italy}
\address[Genova]  {Dipartimento di Ingegneria Biofisica ed Elettronica, \\
                  Universit\`a di Genova, via dell'Opera Pia 11a, I--16145 Genova Italy}

\begin{abstract}
In high-frequency financial data not only returns, but also waiting times between consecutive trades are random variables. Therefore, it is possible to apply continuous-time random walks (CTRWs) as phenomenological models of the high-frequency price dynamics. An empirical analysis performed on the 30 DJIA stocks shows that the waiting-time survival probability for high-frequency data is non-exponential. This fact sets limits for agent-based models of financial markets. 
\end{abstract}

\begin{keyword}
Waiting-time; random walk; statistical finance; efficient market hypothesis
\\ {{\it JEL: \ }} C32;G14
\\ {\it Corresponding author}: Enrico Scalas ({\tt scalas@unipmn.it}), 
\\ url: {\tt www.econophysics.org}
\end{keyword}

\end{frontmatter}

\def\eg{{\it e.g.}\ } \def\ie{{\it i.e.}\ }
\def\sg{\hbox{sign}\,}
\def\sgn{\hbox{sign}\,}
\def\sign{\hbox{sign}\,}
\def\e{\hbox{e}}
\def\exp{\hbox{exp}}
\def\ds{\displaystyle}
\def\dis{\displaystyle}
\def\q{\quad}    \def\qq{\qquad}
\def\lan{\langle}\def\ran{\rangle}
\def\l{\left} \def\r{\right}
\def\lra{\Longleftrightarrow}
\def\arg{\hbox{\rm arg}}
\def\d{\partial}
 \def\dr{\partial r}  \def\dt{\partial t}
\def\dx{\partial x}   \def\dy{\partial y}  \def\dz{\partial z}
\def\rec#1{{1\over{#1}}}
\def\log{\hbox{\rm log}\,}
\def\erf{\hbox{\rm erf}\,}     \def\erfc{\hbox{\rm erfc}\,}
\def\F{\hbox{F}\,}
\def\NN{\hbox{\bf N}}
\def\RR{\hbox{\bf R}}
\def\CC{\hbox{\bf C}}
\def\ZZ{\hbox{\bf Z}}
\def\II{\hbox{\bf I}}


\section{Introduction}

Recently, various authors have discussed the application of
the continuous-time random walk (CTRW) \cite{montroll65} to finance (see Sec. 2 for details). The basic point is that, in financial markets, not only returns, but also waiting times between two consecutive trades can be considered as random variables. To our knowledge, the application of CTRW to economics dates back, at least, to the 1980s. In 1984, Rudolf Hilfer published a book on the application of  stochastic processes to operational planning, where CTRWs were used for sale forecasts \cite{hilfer84}. The (revisited) CTRW formalism has been applied to the high-frequency price dynamics in financial markets by our research group since 2000, in a series of three papers \cite{scalas00,mainardi00,gorenflo01}. Other scholars have recently used this formalism
\cite{masoliver03a,masoliver03b,kutner03}. However, CTRWs have a famous precursor. In 1903, the PhD thesis of Filip Lundberg presented a model for ruin theory of insurance companies, which was further developed by Cram\'er \cite{lundberg03,cramer30}. The underlying stochastic process of the Lundberg-Cram\'er model is a particular example of CTRW.

Among other issues, we have studied the independence between log-returns and waiting times for the 30 Dow-Jones-Industrial-Average (DJIA) stocks traded at the New York Stock Exchange in October 1999. For instance, according to a contingency-table analysis performed on General Electric (GE) prices, the null hypothesis of independence can be rejected with a significance level of 1 \% \cite{raberto02}. In this letter, however, the focus is on the empirical distribution of waiting times.

>From our study, it turns out that the marginal density for waiting times is definitely not an exponential function. After the publication of our paper series, different waiting-time scales have been investigated in different markets by various authors. All these empirical analyses corroborate the waiting-time anomalous behaviour. A study on the waiting times in a contemporary FOREX exchange and in the XIXth century Irish stock market was presented by Sabatelli {\it et al.} \cite{sabatelli02}. They were able to fit the Irish data by means of a Mittag-Leffler function as we did before in a paper on the waiting-time marginal distribution in the German-bund future market \cite{mainardi00}. Kyungsik Kim and Seong-Min Yoon studied the tick dynamical behavior of the bond futures in Korean Futures Exchange(KOFEX) market and found that the survival probability displays a stretched-exponential form \cite{kim03}. Moreover, just to stress the relevance of non-exponential waiting times, !
 a power-law distribution has been
recently detected by T. Kaizoji and M. Kaizoji in analyzing the calm time interval of price changes in the Japanese market \cite{kaizoji03}.

\section{Theory}

The importance of random walks in finance has been known since the seminal
thesis of Bachelier \cite{Bachelier 00} which was completed at the end of the 
XIXth century, more than a hundred years ago. The ideas of Bachelier were further 
carried out by many scholars \cite{Cootner 64,Merton 90}.

The price dynamics in financial markets can be mapped onto a 
random walk whose properties are studied in continuous, rather than
discrete, time \cite{Merton 90}. Here, we shall present this
mapping, pioneered by Bachelier \cite{Bachelier 00}, 
in a rather general way. It is worth mentioning that
this approach is related to those of Clark \cite{Clark 73} and to the
introductory notes in Parkinson's paper \cite{parkinson77}. 
As a further comment, this is a purely phenomenological approach.
No specific assumption on the rationality or
the behaviour of market agents is taken or even necessary.
In particular, it is not necessary to assume the
validity of the efficient market hypothesis \cite{fama70,fama91}.
Nonetheless, as shown below, a phenomenological model can
be useful in order to empirically corroborate or falsify the consequences
of behavioural or other assumptions on markets. 
In its turn, the model itself can be corroborated or falsified by empirical data.

As a matter of fact, there are various ways in which
random walk can be embedded in continuous time. Here, we shall base our approach on the
so-called continuous-time random walk in which time intervals between successive steps are random variables, as discussed by Montroll and Weiss \cite{montroll65}.

Let $S(t)$ denote the price of an asset or the value of an index at time
$t$. In a real market, prices are fixed when demand and offer meet and a 
transaction occurs. In this case,
we say that a trade takes place. Returns rather than 
prices are more convenient. For this reason, we shall take
into account 
the variable $x(t) = \log S(t)$, that is the logarithm of the price.
Indeed, for a small price variation $\Delta S = S(t_{i+1}) - S(t_{i})$, the 
return $r = \Delta S/S(t_{i})$ and the logarithmic return 
$r_{log} = log[S(t_{i+1}) / S(t_{i})]$ virtually coincide. 

As we mentioned before, 
in financial markets, not only prices can be modelled as random variables, 
but also waiting times between two consecutive 
transactions vary in a stochastic fashion. 
Therefore, the time series $\{ x(t_i) \}$
is characterised by $\varphi(\xi, \tau)$, the 
{\em joint probability density}
of log-returns $\xi_{i} = x(t_{i+1}) - x(t_{i})$ and of waiting times 
$\tau_i = t_{i+1} - t_{i}$. The joint density
satisfies the normalization condition 
$\int \int d \xi d \tau \varphi (\xi, \tau) = 1$.

Montroll and Weiss \cite{montroll65} have shown that the Fourier-Laplace
transform
of $p(x,t)$, the probability density function, {\em pdf}, of
finding 
the value $x$ of the price logarithm (which is the diffusing quantity in 
our case) at time $t$, is given by:
\begin{equation}
\label{montroll}
\widetilde{\widehat p}(\kappa, s) = {1-\widetilde \psi(s) \over s}\,
{1 \over 1- \widetilde{\widehat \varphi}(\kappa, s)}\,,
\end{equation}
where 
\begin{equation}
\label{transform}
\widetilde{\widehat p}(\kappa, s) = 
\int_{0}^{+ \infty} dt \; \int_{- \infty}^{+ \infty} dx \, 
\e^{\,\ds -st+i \kappa x} \, p(x,t)\,,
\end{equation}
and $\psi(\tau) = \int d \xi \, \varphi(\xi, \tau)$ 
is the waiting time pdf.

The space-time version of eq. (\ref{montroll}) can be derived by probabilistic considerations \cite{mainardi00}. The following integral equation gives the probability density, $p(x,t)$, for the walker being in position $x$ at time $t$, conditioned by the fact that it was in position $x=0$ at time $t=0$:
\begin{equation}
\label{masterequation}
p(x,t) =  \delta (x)\, \Psi(t) +
   \int_0^t \, 
 \int_{-\infty}^{+\infty}  \varphi(x-x',t-t')\, p(x',t')\, dt'\,dx',
\end{equation}
where $\Psi(\tau)$ is the so-called survival function. $\Psi(\tau)$ is related to the marginal waiting-time probability density $\psi(\tau)$. The survival function $\Psi(\tau)$ is:
\begin{equation}
\label{survival}
\Psi(\tau) = 1 - \int_{0}^{\tau} \psi (\tau') \, d \tau' = \int_{\tau}^{\infty} \psi (\tau') \, d \tau'.
\end{equation}

The CTRW model can be useful in applications such as speculative option pricing by Monte Carlo simulations or portfolio selection. This will be the subject of a forthcoming paper.

\section{Empirical evidence}

In Fig. 1, the waiting-time complementary cumulative distribution function (or survival function) $\Psi(\tau) = 1- \int_{0}^{\tau} \psi(t') dt'$ is plotted for three different periods of the day (morning 9:00 - 10:59; midday 11:00 - 13:59, afternoon 14:00 - 17:00) and for the GE time series of October 1999. In the above formula, $\psi(\tau)$ represents the marginal waiting-time probability density function. $\Psi(\tau)$ gives the
probability that the waiting time between two consecutive trades is greater than the given $\tau$.
The lines are the corresponding standard exponential complementary cumulative distribution functions: $\Psi(\tau) = \exp(-\tau/\tau_0)$, where $\tau_0$ is the empirical average waiting time. An eye inspection already shows the deviation of the real distribution from the exponential distribution. This fact is corroborated by the Anderson-Darling test \cite{stephens74}. The Anderson-Darling (AD) $A^2$ values for the three daily periods are, respectively, 352, 285, and 446. Therefore, the null hypothesis of exponential distribution can be rejected at the 5 \% significance level.

In Tab. 1, the values of the AD $A^2$ statistics are given for the 30 DJIA stocks traded in October 1999. In all these cases the null hypothesis of exponentiality can be rejected at the 5 \% significance level.

\section{Conclusions and discussion}

Why should we care about these empirical findings on the waiting-time distribution? This has to do both with the market price formation mechanisms and with the bid-ask process. {\it A priori}, one could argue that there is no strong reason for independent market investors to place buy and sell orders in a time-correlated way. This leads to a Poisson process with exponential survival function for the bid-ask process. If price formation were a simple thinning of the bid-ask process, then exponential waiting times should be expected between consecutive trades as well \cite{cox79}. Eventually, even if empirical analyses should show that time correlations are already present at the bid-ask level, it would be interesting to understand why they are there. In other words, the empirical results on the survival probability set limits on statistical market models for price formation.

\section*{Acknowledgements}

We would like to acknowledge useful discussions with Sergio Focardi of The Intertek Group. This work has partially been carried out in the frame of the INTAS project 00-0847.

\begin{figure}
\includegraphics{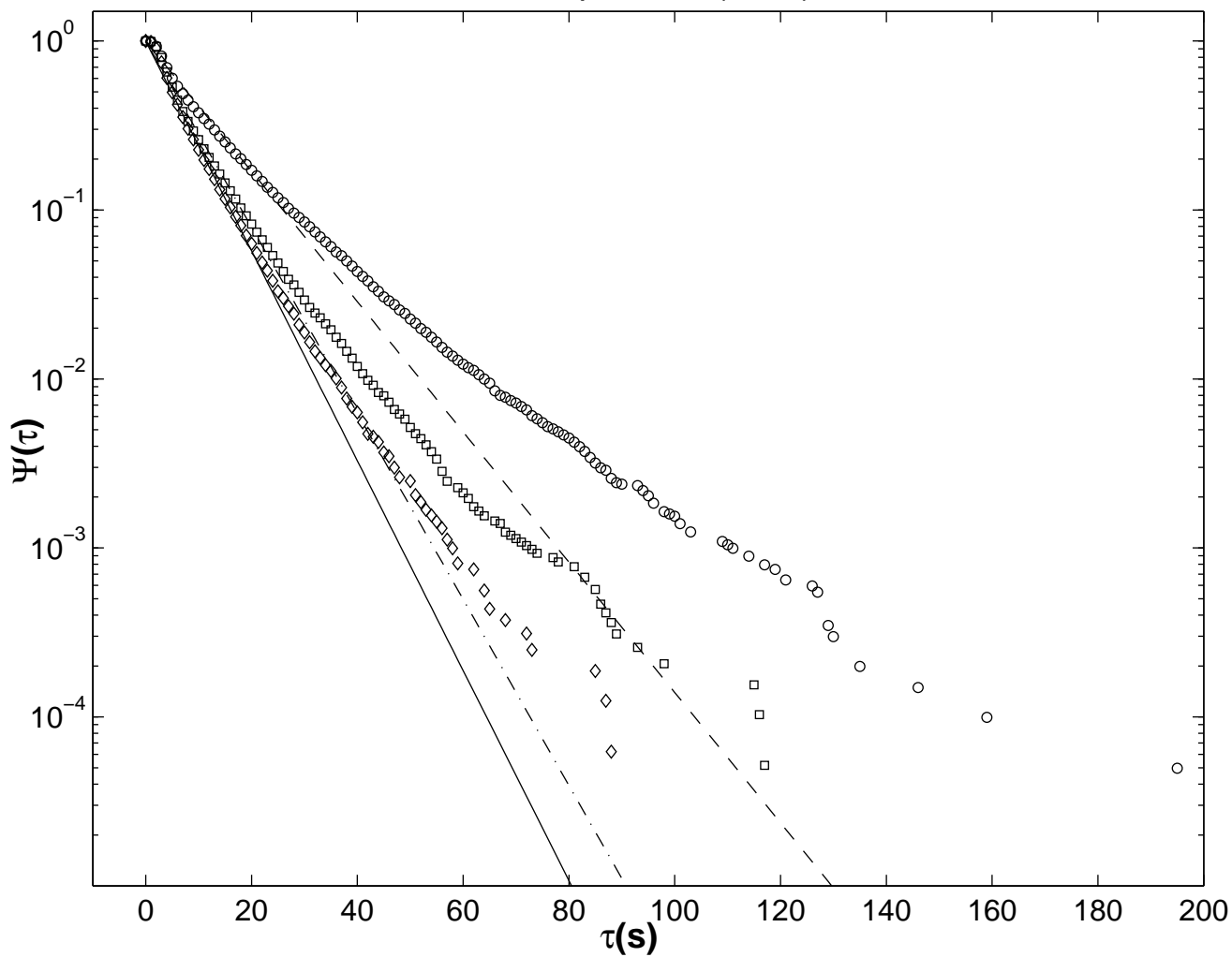}
\caption{\label{fig:epsart} Waiting-time complementary cumulative distribution function $\Psi(\tau)$ for GE trades quoted at NYSE in October 1999. Open diamonds represent $\Psi(\tau)$ for the morning hours (9:00 -- 10:59). There were 16063 trades in this period in October 1999. The solid line is the corresponding standard exponential complementary cumulative distribution function with $\tau_0 = 7.0$ s. Open circles represent $\Psi(\tau)$ for the period around midday (11:00 -- 13:59). There were 20214 trades in this period in October 1999. The dashed line is the corresponding standard exponential complementary cumulative distribution function with $\tau_0 = 11.3$ s. Open squares represent $\Psi(\tau)$ for the afternoon hours (14:00 --  17:00). There were 19372 trades in this period in October 1999. The dash-dotted line is the corresponding standard exponential complementary cumulative distribution function with $\tau_0 = 7.9$ s. The day was divided into three periods to evide!
 nce seasonalities.}
\end{figure}

\begin{table}
\begin{center}
\begin{tabular}{|c|c|c|c||c|c|c|}
\hline
Stock&$\tau_{0}^{mo} (s)
$&$\tau_{0}^{mi} (s)
$&$\tau_{0}^{af} (s) $&$A^{2}(mo)$&$A^{2}(mi)$&$A^{2}(af)$ \\
\hline\hline
AA&27.1&40.0&28.8&29.2&66.0&44.8\\
\hline
ALD&21.2&30.8&23.4&21.8&55.5&33.8\\
\hline
AXP&11.8&18.5&11.7&81.7&102.5&130.7\\
\hline
BA&22.0&32.0&22.6&17.4&20.2&21.2\\
\hline
C&7.1&10.5&8.2&252.2&142.8&210.7\\
\hline
CAT&29.2&42.4&31.6&72.3&128.7&64.6\\
\hline
CHV&22.1&34.3&27.1&104.4&121.5&64.9\\
\hline
DD&20.3&30.8&22.1&22.9&44.3&36.1\\
\hline
DIS&15.2&20.8&16.6&53.4&53.4&74.7\\
\hline
EK&34.1&51.2&36.3&24.8&34.8&44.3\\
\hline
GE&7.0&11.3&7.9&351.9&284.7&445.6\\
\hline
GM&24.6&36.6&27.0&22.4&60.8&40.9\\
\hline
GT&34.3&55.5&37.9&73.7&95.7&54.1\\
\hline
HWP&10.4&16.1&12.7&94.8&77.8&100.8\\
\hline
IBM&8.9&10.0&9.2&409.6&472.5&489.5\\
\hline
IP&24.8&36.3&27.0&25.0&37.2&19.4\\
\hline
JNJ&16.1&23.0&17.7&30.4&35.6&38.0\\
\hline
JPM&17.0&29.5&19.0&33.0&85.2&85.8\\
\hline
KO&12.9&18.3&14.4&44.5&37.8&44.1\\
\hline
MCD&19.4&29.3&22.1&40.9&72.7&44.1\\
\hline
MMM&30.1&42.0&30.4&80.1&86.8&37.5\\
\hline
MO&11.4&15.6&12.9&74.2&89.0&75.2\\
\hline
MRK&11.7&16.8&13.2&133.1&136.0&189.8\\
\hline
PG&16.2&23.6&17.9&43.5&37.2&48.8\\
\hline
S&23.4&38.8&28.6&40.1&23.0&41.6\\
\hline
T&8.8&12.2&10.6&193.2&179.1&208.9\\
\hline
UK&40.4&69.1&46.7&33.8&72.4&47.2\\
\hline
UTX&28.5&39.3&29.0&33.7&62.9&58.0\\
\hline
WMT&12.5&18.2&14.9&105.2&110.6&139.1\\
\hline
XON&12.0&19.6&14.1&104.8&121.4&129.0\\
\hline
\end{tabular}
\end{center}
\caption{For each daily period, the table gives the values of the empirical average waiting time $\tau_{0}$ and the AD statistics $A^{2}$
\cite{stephens74}.}\label{ADtest}
\end{table}

\end{document}